# Ageing as a price of cooperation and complexity

## Self-organization of complex systems causes the gradual deterioration of constituent networks

Huba J. M. Kiss[1], Ágoston Mihalik[1], Tibor Nánási[1], Bálint Őry[1], Zoltán Spiró[1], Csaba Sőti[1] and Peter Csermely[1]*

**The network concept is increasingly used for the description of complex systems. Here we summarize key aspects of the evolvability and robustness of the hierarchical network-set of macromolecules, cells, organisms, and ecosystems. Listing the costs and benefits of cooperation as a necessary behaviour to build this network hierarchy, we outline the major hypothesis of the paper: the emergence of hierarchical complexity needs cooperation leading to the ageing (i.e. gradual deterioration) of the constituent networks. A stable environment develops cooperation leading to over-optimization, and forming an 'always-old' network, which accumulates damage, and dies in an apoptosis-like process. A rapidly changing environment develops competition forming a 'forever-young' network, which may suffer an occasional over-perturbation exhausting system-resources, and causing death in a necrosis-like process. Giving a number of examples we demonstrate how cooperation evokes the gradual accumulation of damage typical to ageing. Finally, we show how various forms of cooperation and consequent ageing emerge as key elements in all major steps of evolution from the formation of protocells to the establishment of the globalized, modern human society.**

**Introduction: Evolvability, robustness and ageing of hierarchical networks**
The network approach proved to be a highly efficient tool to describe various levels of the hierarchical organization of complex systems from macromolecules to the currently emerging world-wide social networks. The network description needs the identification of separable subsets of the system as network elements, and a catalogue of their interactions as network contacts or links. In most of the cases network elements themselves can be perceived as networks. Thus, elements of social networks, human individuals are networks of organs and cells, cells are networks of proteins, and proteins are networks of amino acids. Networks display a lot of rather general properties, such as
(a) *small-worldness*, meaning the existence of short pathways between most network elements;
(b) the *existence of hubs*, which have a much higher number of neighbours than the average;
(c) a *modular structure*, which organizes networks to various overlapping groups;
(d) the co-existence of *strong and weak links*, where the link-strength is usually defined as the real, physical strength of the connection, or as the probability of interactions and
(e) the existence of a *network skeleton*, which is the subset of most important pathways in the network.

Regretfully, the scope of the current paper does not allow us to give exact definitions and a detailed description of all these network properties, therefore, the reader is referred to a number of recent reviews for more details.[1–4]

In the previous list of network properties almost all general network features referred to a structural, topological description of networks. However, an even more important task is the characterization of network dynamics, which became a centrepiece of network studies in the last couple of years. Networks continuously accommodate novel members, lose their original elements, as well as build, erase and rearrange their links. Networks may even undergo a profound structural reorganization, called topological phase transition, when they experience a large change in the resources providing the energy to maintain their links, or suffer a large stress, i.e. an abrupt change in the number and magnitude of perturbations arriving from the network environment and disturbing their original structure.[1–7]

[1]Department of Medical Chemistry, Semmelweis University, P O Box 260., H-1444 Budapest 8, Hungary.
Funding agency: Work in the authors' laboratory was supported by the EU (FP6-518230) and the Hungarian National Science Foundation (OTKA K69105). C.S. is a Bolyai Research Scholar of the Hungarian Academy of Sciences.
*Corresponding author: Peter Csermely, Department of Medical Chemistry, Semmelweis University, P. O. Box 260., H-1444 Budapest 8, Hungary. E-mail: csermely@eok.sote.hu

Structural changes of biological networks often respond to the novel stimulus of the environment, especially if the stimulus is repeatedly experienced. This adaptation of network structure to the environmental signals – which is similar to the training of artificial neural networks – can be perceived as a learning process. The plasticity of a network enabling these adaptive changes can be described as network evolvability. In its more restricted and original sense, the evolvability of a biological system is the capacity of the system to generate a heritable phenotypic variation.[2,8,9] Evolvability is by itself an inheritable property,[10,11] which shows that the plasticity of biological networks is a carefully regulated key feature of the accommodation to the magnitude, speed, rhythm and unexpectedness of environmental changes.

The evolvability of a network (i.e. in this, broader sense its potential to produce and accommodate innovations changing the properties of the complex system) is balanced by network robustness. Robustness and evolvability at the level of network structure display a rather antagonistic relationship. In stark contrast with structural robustness, robustness at the level of network function (i.e. at the level of the phenotype of the complex system encoded by the network, Box 1) does not hinder, but actually may even promote the evolvability of the network structure.[12] Recent studies show, that variable network structures displaying a robust phenotype can step-by-step access large amounts of structural variation avoiding the penalty of natural selection.[13,14] However, there is an intricate balance in biological systems to maintain their functional robustness, and to allow the chance of an evolutionary change by preserving their evolvability. The interplay of evolvability-mediated innovations and the buffering effects of functional robustness often leads to jumps between stability-islands causing a bi-stable or biphasic network behaviour, where two markedly different functional states of the network are much more populated than any of the myriads of the potential states between them.[15]

> **Box 1. Phenotypic, functional robustness of biological networks.** Robustness of biological networks at the level of network function (i.e. at the level of the phenotype of the complex system described by the network) is helped by a number of mechanisms.
> 1. **Strong links** (meaning intensive, high probability, high affinity interactions) often form negative or positive feedbacks helping the biological system to return to the original state (attractor) or jump to another, respectively. This systems control enables the system to move between two stable states.
> 2. The contribution of **weak links** (meaning non-intensive, low probability, low affinity interactions) is more diffuse. Weak links provide (i) alternative, redundant, degenerate pathways; (ii) flexible connections disjoining network modules to block perturbations and re-assembling the modules in a slightly altered fashion, and (iii) additional, yet unknown mechanisms buffering the effects of the original perturbation further, and de-coupling physical perturbations from functional activities. We must note that weak links grossly outnumber strong links in cellular networks, which may often make the effects of their fail-safe mechanisms larger than those of negative or positive feedbacks.
> 3. Finally, robustness of cellular networks is also helped by an **increased average robustness of their elements** (e.g. proteins), which, most of the time are networks by themselves.[2,12]

In this paper we consider ageing from the network sense.[16–18] We define ageing as the gradual phenotypic appearance of the consequences of the suboptimal conditions caused by the suboptimal distribution of available resources of the complex system between its sub-systems. We list the most common consequences of suboptimal sub-system conditions below:
- decline of costly repair systems;
- accumulation of damage;
- increase in noise;
- decrease in complexity;
- deterioration of network plasticity;
- loss of adaptation potential.

Due to the (1) plethora of possible segregations of the various sub-systems; (2) the numerous solutions of suboptimal resource distribution between sub-systems, and (3) the improbably high number of possible scenarios how the suboptimal resource distribution will gradually affect sub-system functions – ageing has been increasingly perceived as one of the most complex phenomena of biological systems. Various ageing theories exist, which highlight one or another type of suboptimal resource distribution as a cause of the ageing process.

- According to the antagonistic pleiotropy theory of ageing, genes, which are preferable during early development, become detrimental at later stages. This shows a suboptimal resource distribution at various stages of the life-cycle.
- The disposable soma theory of ageing highlights the relocation of resources from somatic maintenance towards increased fertility.
- The "network theory of ageing" describes the balance between various types of damage and repair mechanisms in the special case of the ageing of cellular systems accommodating a multitude of suboptimal resource distribution scenarios.[16–19]

From the general approach of sub-optimal resource distribution as a major cause of ageing it is clear that ageing (in this network sense) will not occur, if the system enjoys unlimited resources. However, this is a rather unlikely scenario at its extreme form. A good compromise allowing a close-to-optimal function and maintenance of all sub-systems may be regarded as a form of successful, healthy ageing, which avoids most of the accumulating damage caused by the usual, sub-optimal resource distribution.

In the following sections of the paper we will show that cooperation of the constituent networks is necessary to build up the hierarchical network structure from macromolecules to the globalized social network. We will describe the costs of this beneficial cooperation in a number of examples. We then outline our central hypothesis that cooperation leads to an additional, previously unrecognized cost: the gradual deterioration of subsystems typical to the ageing process, and finally we will show the generality of this statement by pinpointing potential ageing phenomena of the constituent networks at each major steps of the evolution[20] starting in the formation of the protocell, through the development of coordinated replication, transcription and protein synthesis, the establishment of the eukaryotic cell, sexual proliferation, differentiated multicellular organisms, animal social groups and finally, human societies.

**Cooperation: costs, benefits and the emergence of hierarchical complexity**
Cooperation is a joint action for a common goal, which requires information exchange and strategy adjustment of the participating partners. Development of cooperation between the constituent networks is a key step in many major transitions in evolution.[20] Game theory describes a number of biologically relevant examples, where cooperation provides a smaller income for all participating agents than the opposing behaviour, cheating, or defection. However, in a large subset of these games, called social dilemmas, the limited private income of cooperators, i.e. the cost of cooperation is paralleled with the maximization of overall income for the whole cooperating community. How can selfish replicators forgo a part of their own replication potential, and use their resources to help each other? Various examples of repeated games restricted to neighbours in a network context show that several highly interrelated mechanisms help the survival of cooperation: e.g. kin-selection; direct reciprocity; indirect reciprocity; network reciprocity or group selection.[21–24]

Let us illustrate the costs of cooperation on the rather extreme examples of self-destructive cooperation. Here cooperation has an especially high price: the loss of reproduction or death of the cooperating individual upon offering a higher accessibility of public goods for the overall population. Obviously, the sacrifice of self-destructive cooperation can be typical only to a limited fragment of the whole population, since otherwise it would lead to a general extinction.[25] A number of examples of self-destructive cooperation from bacterial lysis to examples of human behaviour are listed in Box 2.[26–33] Interestingly, even general, network-related benefits of cooperative behaviour can be perceived as special forms of self-destructive cooperation. Cooperators must form cooperation-enriched islands to survive, and the existence of hubs may also overcome the temptation to defect[21–24,34,35]. If we consider the costs of maintaining a network contact, both the emergence of densely connected islands and – especially – hubs (i.e. network marvels, who have countless numbers of acquaintances) may also be perceived as a rather costly, and in part, even self-destructive act, since these individuals certainly have less energy and time for their own offspring.

> **Box 2. Examples of self-destructive cooperation**
> 1. Bacterial virulence factors leading to destruction by the innate immune system
> - *Salmonella typhimurium* Type III secretion systems and flagella enhancing gut inflammation and colonization[26]
> 2. Bacterial lysis
> - *Streptococcus pneumoniae* pneumolysin helping lung colonization[27]
> - *Clostridium difficile* TcdA enhancing gut inflammation and colonization[28]
> 3. Bacterial quorum sensing leading to reduction of virulence
> - *Pseudomonas aeruginosa* LasR quorum sensing regulator enhancing survival in dense populations[29,30]
> 5. Yeast invertase enzyme secretion
> - *Saccharomyces cerevisiae* Suc2 invertase secretion enhancing the availability of glucose[31]
> 6. Hara-kiri of neutrophil granulocytes
> - neutrophils kill themselves by releasing extracellular structures of chromatin and granule proteins called neutrophil extracellular traps (NETs) to capture and kill invading bacteria[32]
> 7. Animal infertility
> - workers of insect colonies enhancing offspring survival
> 8. Alarm calls
> - several animals, e.g. marmots, risk self-sacrifice warning other members of their community on the appearance of a predator[33]
> 9. Soldiers' heroism
> 10. Altruism of nuns, monks and catholic priests
> 11. Blood donation
> 12. Charity donations
>
> In many of the above examples the benefit is not evident at the level of the immediate community of the self-destructive cooperator, but becomes manifest at one or more levels higher in the hierarchical complexity of networks. As an additional example of these multi-layer effects, we may consider antibodies 'altruistic', when they sacrifice themselves co-degrading with the antigen for the health of the hosting organism living at least two levels higher in the hierarchical complexity.

Cooperation was necessary to build the growing layers of embedded network hierarchy including the emergence of eukaryotic cells, multicellularity and eusociality.[20] Recently Richard E. Michod[36] developed a plausible model to show that in case of a hierarchical network, cooperation transposes fitness from the lower level (e.g. from the costs of the individual cells) to the higher level (e.g. to the benefits of the multicellular assembly) and in this way it may extend the 'fitness-window' offering novel means of survival. In these non-zero-sum games cooperation acts as a mediator of conflict reducing the selection inside the organism or group, and increasing the selection between organisms. A similar effect has been mentioned before as group-selection.[23] The recent work of Efferson et al.[37] demonstrates the same transposure of fitness from the individuals to their social group showing that cultural processes (e.g. the introduction of discriminating markers for group identification) can favour the evolution of previously unfavourable behavioural traits, like altruistic cooperation. We will show the consequences of such fitness-transposure scenarios to the network structure and cooperation behaviour in the later section describing "ageing as a consequence of cooperation".

The importance of cooperation between the elements of growing layers of the hierarchical complexity of biological systems can be perceived as an effort of the self-organizing complex systems to become more autonomous by building in an increased amount of complexity and flexibility,[38] as well as to stabilize as large segment of their environment as possible. An example of this latter behaviour is the network-building property of horizontally transferred genes in the host's regulatory network.[39] However, the preservation of the evolvability of the system requires the preservation of non-cooperating, creative network elements as sources of innovation and network plasticity. These creative elements are exemplified by active centres of enzymes, molecular chaperones of cells, stem cells of complex organisms, or creative persons of social networks. Creative elements perform a random sampling of the whole network, and provide flexible links between variable network communities. The continuous jumps of creative elements exclude their prolonged cooperation with any of the constituent network groups.[34,40]

**Two evolutionary strategies of survival: co-occurrence of cooperation and ageing**
In the following two sections we outline our major hypothesis arguing that cooperation has a more general cost besides those of self-destructive cooperation mentioned before: ageing in the network sense of gradual deterioration of the cooperating units. First, we summarize two major evolutionary strategies of survival, called small and big phenotypes at the level of various individuals,[41] or r/K-strategies at the level of ecosystems.[42] We show that in the slow and efficient small phenotypes (K-strategists) cooperative behaviour is leading to a prolonged accumulation of damage typical to the ageing process. Continuing the examples of simultaneous ageing and cooperation we show that anti-ageing homologues are significantly enriched among the 'non-cooperative/cheater genes' of the amoeba, *Dictyostelium discoideum*.[43]

As we mentioned before, complex systems often show biphasic behaviour. Our first example of cooperation-related ageing is the small phenotype of the small and big phenotype-pair of humans and a wide range of other organisms summarized recently by Bateson et al.[41] In this dual strategy, small phenotypes with small size and slow metabolism become accommodated to adverse circumstances, while big phenotypes may enjoy a larger size and a rapid metabolism due to more abundant resources. With a large individual variation, small phenotypes have a small number of offspring, cooperate in their nursing, and have a long life expectancy. Here ageing is a typical process. On the contrary, individuals of the big phenotype have a larger number of offspring, and are competitive. Here ageing is not a typical process. Caloric restriction has been shown to prolong life giving a chance of a typical ageing process. Caloric restriction also induces a reduction of competition-related fertility, and closely resembles the transition of the big → small phenotype behaviour.[44,45] However, this transition is not immediate, but may require as many as three generations for completion.[41] Ageing of the small phenotypes might be a simple consequence of their longer life, since typical, deteriorating signs of ageing may not simply occur during a shorter lifespan. However, we hypothesize that in the resource-restricted environment of small phenotypes the cooperation-based sub-optimal resource allocation gives suboptimal, degradation-prone conditions for many individuals, while big phenotypes may enjoy the abundance of resources without self- or community-imposed limitations.

A similar dual strategy can also be observed at one level higher in the network hierarchy, in the ecosystems. The well-known r/K selection theory of ecology describes the emergence of two types of individuals: (1) r-strategists produce a lot of offspring, each of which has a relatively low probability of surviving to adulthood, and occupies every available niches, while (2) K-strategists live in crowded niches investing a lot to their relatively few offspring, which have a high probability to survive to adulthood. K-strategists often enjoy long life expectancies and populate relatively stable and predictable environments.[42] K-strategists tightly cooperate within their group, e.g. family. On the contrary, r-strategists competitively explore the whole environment. Exponentially growing bacteria, yeast and stem cells are typical r-strategists, where cooperation is not prevalent, and the small life-expectancy does not allow the accumulation of massive age-related damage. Moreover, yeast and stem cell r-strategists often 'shed off' damaged proteins and other cellular components by asymmetric cell division, which segregates most of the damaged material to mother-cells, producing a 'fresh' daughter-cell as an offspring.[46,47] r-strategists often die non-predictably and abruptly, and do not age in the predictable and long sense of the phenomenon. On the contrary, K-strategists, while enjoying a long life and cooperating to protect and raise their offspring, accumulate all the damage to the extent r-strategists never reach. Moreover, K-strategists have to use the limited resources very efficiently, where again cooperation brings self- or community-imposed limitations causing suboptimal, degradation-prone conditions for many individuals, while r-strategists competitively explore, and inefficiently use the abundance of resources. Similarly to the transitions of the big and small phenotypes mentioned above, the ratio of r- and K-strategies may also change according to the environment. Moreover, many organisms may also display an intermediate state. As an example of this, mice and other rodents can mostly be categorized as r-strategists, while showing a lot of aspects of cooperation and ageing.

An interesting example of the r → K-strategy change is the change in the soil microbial community after meadow mowing. Mown-meadow allows the growth of competing r-strategist microbes, due to the enriched resources by the smaller consumption of the mown grass. On the contrary, un-mown-meadow has a predominance of cooperating K-strategist microbes due to the intensified competition with the surrounding grass for resources.[48] The mown → un-mown; r- → K-strategy transition resembles well to the transition of caloric restriction – observed now at the community level.

**Table 1. General properties of the two major evolutionary strategies of survival: the association of cooperation and ageing**

| Property | Global, loose cooperation (big phenotype, r-strategist) | Local, tight cooperation (small phenotype, K-strategist) |
|---|---|---|
| Level of cooperation | low | high |
| Community-control | loose | tight |
| Fluctuation, noise, creativity | high | low |
| Ageing | atypical, fast | typical, long |
| Death | unpredictable | predictable |
| Efficiency of solving simple, goal-oriented tasks | low | high |
| Sustainability in a changing environment | high | low |
| Status and fate in a stable environment with a low intensity of input, resources | gradual, stochastic disintegration, fast ageing, and unpredictable death | optimal |
| Status and fate in an unstable environment with a high intensity of input, resources | optimal | accumulation of damage, long ageing, leading to a predictable death |
| **Summary of strategy** | 'forever young' | 'always old' |

We summarize the major properties of the two dualities of small and big individuals as well as of r- and K-strategists[41,42] in Table 1. The competitive strategy may appear as a big phenotype at the level of the individual, and as an r-strategist at the level of the whole ecosystem. Conversely, the cooperative strategy usually appears as a small phenotype at the level of the individual, and as a K-strategist at the level of the ecosystem. Competition is global and the resulting network is loose due to the speedy proliferation and network expansion of big/r-strategists. On the contrary, cooperation is local and the resulting network is tight due to the slow, restricted, sparing life of small/K-strategists. The local character of cooperation is also in agreement with the preference of cooperating, local islands as described before.[21,24,34]

As a summary of this comparison, competing big individuals, or r-strategists remain fast growing, expanding, inefficient 'forever-young' organisms, and achieve their success in this way. On the contrary, cooperating small individuals, or K-strategists behave as slow, wise, farseeing, efficient, 'always-old' organisms, and achieve their success in this way.

As we noted before, most of the times none of the above strategies exists in its pure form. In agreement with the general picture outlined before, the transition between the global competition → local cooperation resembles to the previously mentioned topological phase transition of networks occurring at the reduction of resources, i.e. in times of prolonged stress.[5–7] As an extension of this analogy, in simple game models environmental stress promoted cooperation even under circumstances, when cooperative behaviour was too costly and was avoided in normal conditions.[49]

**Co-occurrence of cheating and anti-ageing genes**
To provide a second example showing the correlation of ageing with cooperative behaviour we analyzed the recently described *Dictyostelium discoideum* model, where a rich genetic background of asocial, defecting, cheater (or cooperating, loser) mutants was uncovered allowing the hosting amoebas to produce more (or less) than their fair shares of spores in mixed colonies.[43] We have searched for the homologues of these genes and determined, if these homologous genes had been demonstrated to play any role in the ageing process. The results are summarized in Table 2. We found six cheater genes, the synaptogenic Unc-10,[50] the stem cell- and NMDA receptor-related Nfyc/NARG2,[51,52] a member of the *old-1/old-2* tyrosine kinase family,[53] the Ca-calmodulin activated kinase kinase 2 (CAMKK2),[54] the ubiquitin-conjugating enzyme 2 and the mitochondrial fission-related chondrocyte protein with a poly-proline region (CHPPR),[55] which all participate in various processes hindering the ageing process. Notably, the best documented case of the six 'anti-ageing' genes, the Unc-10 homologue, was among the 6 strongest, 'conditional' cheaters, which produced more spores under the experimental conditions of the test applied in the original paper.[43] Six out of the tested 31 'cheater' and 6 'loser' genes may not seem to be a large hit-rate. However, we would like to note that in the unlisted 31 cases the lack of identified homologues (11 cases) or the lack of available biological information was rather predominant. Moreover, we did not find a single example, where the contrary of the expectations was ever described. We believe that this uneven occurrence of data gives a rather strong support to the note that the genetic background of the non-cooperating phenotype is helpful to reduce the deleterious consequences of ageing. This assumption is further substantiated by the fact that screening of the residual 161 genes examined in ref. 43 resulted in 6 additional genes related to ageing or longevity listed in the legend of Table 2. When identifying these genes we took all possibilities into account including their participation in neurodegenerative diseases, which is only indirectly related to the ageing process. In spite of this rather generous sampling the difference between the likelihood to find a 6-gene cohort in the 31, or 161 gene-samples is highly significant ($p < 0.005$ using the chi-square probe), which shows that ageing-related genes are significantly enriched among the cheater genes of *Dictyostelium discoideum*. In summary, these data provide an additional piece of corroborative evidence that ageing is a price for cooperation.

**Similarity of cooperating and aged network structures**
As a third and more general example, on Figure 1 we illustrate the typical network structure of both a competitive and a cooperative network. We highlight the main structural features of the two systems in the followings.

Globally competitive, loose systems (Figure 1A)
- have a looser network structure with a large number of predominantly weak links
- are less integrated locally, but more integrated globally
- have large overlaps of their modules, and
- have a suppressed importance of their network skeleton.

On the contrary, local, tightly cooperating systems (Figure 1B)
- have a tight local structure with a small number of predominantly strong links
- are more integrated locally, but less integrated globally
- have small overlaps of their modules, and
- have a key importance of their network skeleton (for details, see Suppl. Table 1 of ref. 34).

It is interesting to note that competitive networks resemble to the stratus-type, or 'stringy-periphery' networks, while cooperative networks are similar to the cumulus-type, or 'multi-star' networks.[56,57]

**Table 2. Ageing/longevity-related homologues of asocial (non-cooperating, cheater, defecting) genes in the amoeba *Dictyostelium discoideum***

| Asocial amoeba gene | Level of asociality | Ageing-related homologous gene(s) | Relation to ageing/longevity |
|---|---|---|---|
| DDB0219502 | 64.7% | Unc-10 (*C. elegans*) | causes an approx. 35% increase in longevity[50] |
| DDB0191265 | 63.2% | Nfyc (*C. elegans*); NARG2 (human) | involved in developmental processes in stem cells and neurons[51,52] |
| DDB0191503 | 62.3% | Src-2 (*C. elegans*) | member of longevity inducing *old-1/old-2* tyrosine kinase family[53] |
| DDB0220010 | 56.8% | CAMKK2 (human) | preservation of synaptic plasticity in ageing[54] |
| DDB0187308 | 56.0% | Ubiquitin conjugating enzyme E2 | participation in proteasomal degradation of damaged proteins |
| DDB0169123 | 52.2% | CHPPR (human) | participation in antioxidation defence[55] |

The Table summarizes those genes, which influenced the social behaviour of the amoeba *Dictyostelium discoideum*,[43] and had a homologous gene in the database of GenAge[77] or homologous ageing-related gene or protein in Pubmed. Gene homology has been established by the help of InParanoid[78], Wormbase[79] and Ensembl[80] databases. From the residual 161 *Dictyostelium discoideum* genes examined in ref. 43, we have found 6, which were related to either longevity, or accelerated ageing (in parentheses please find the PubMed ID of the respective papers, where appropriate): acad8 (PMID: 17387528); DDB0231250 chaperonin; irlB (PMID: 11846374); kynureine aminotransferase (PMID: 7650530); psmD1 proteasomal subunit and rpl10 (PMID: 17174052).

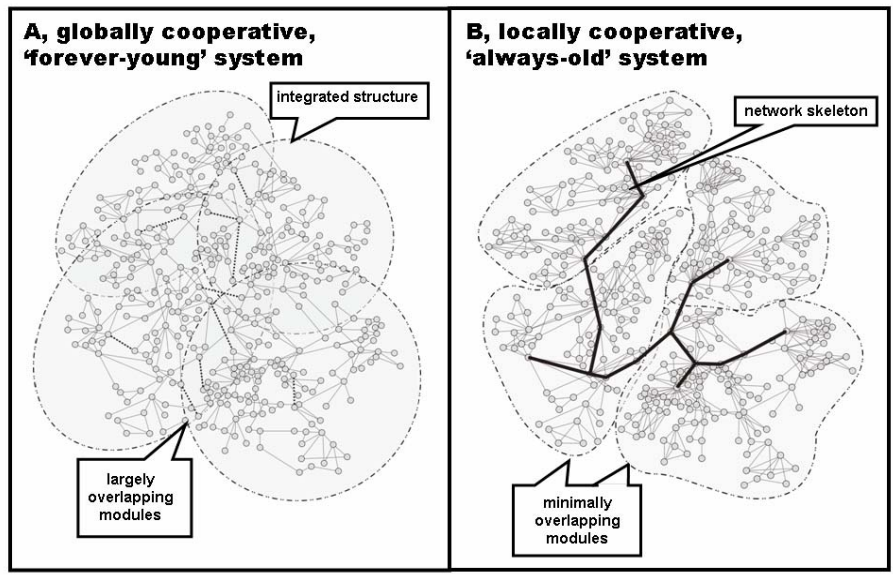

**Figure 1. Typical network structure of globally, loosely cooperative, locally competitive, 'forever-young' and locally, tightly cooperative, 'always-old' systems. A:** Networks of globally cooperative, locally competitive systems have a looser structure with a large number of predominantly weak links, are less integrated locally but more integrated globally, have large overlaps of their modules, and have a suppressed importance of their network skeleton. This structure is typical to the exploratory, 'forever-young' systems. **B:** Networks of locally cooperative systems have a tight local structure with a small number of predominantly strong links, are more integrated locally but less integrated globally, have small overlaps of their modules, and have a key importance of their network skeleton. This structure is typical to the restrictive, 'always-old' systems.

The increased importance of network skeleton in cooperating, ageing networks has another major consequence. So far throughout the paper we were concentrating to protein-protein interaction networks at the cellular level. In protein-protein interaction networks the network skeleton, i.e. a set of network pathways, where most information travels in the network, may well be identical with the signalling network.[58] Signalling networks, especially insulin signalling, play a predominant role in ageing.[59] Interestingly, insulin signalling is situated in the intersection of many signalling pathways[60,61] – forming a kind of 'skeleton of the skeleton' in the global network context. Thus, the importance of the skeleton of cooperating networks agrees well with the well-known importance of signalling in the ageing process.

An ageing network makes the older nodes rather isolated, and allows only a limited local spread of information instead of global coupling. The limited availability of resources in an ageing network reduces the number of links, which diminishes modular overlaps, and gives and increased value of the remaining, tighter local structures and network skeleton.[18,62] These age-induced changes of network structure closely resemble the structure of locally cooperative, tight networks, which shows that the two processes, the emergence of continuous, unchanged local cooperation, and the age-induced adaptation go hand-in-hand, and enhance each other in network evolution.

**Ageing as a consequence of cooperation**
In the previous section we demonstrated the co-occurrence of cooperation and ageing in many examples, including the duality of cooperative and ageing evolutionary strategies; the enrichment of the anti-ageing gene homologues of the non-cooperating genes of *Dictyostelium discoideum*; and the description of the similarities of cooperative and aged network structures. Here, we extend our hypothesis showing that cooperation not only accompanies but actually leads to ageing, both at the level of the network, and at one level higher, where the original network is only one of the cooperating partners.

(1) *At the level of the original network* cooperation leads to the specialization of elements. Here the optimal equilibrium of repair and damage may be disturbed by over-optimization, where repair becomes predominant over damage. Network elements form a rigid, locally integrated network typical to aged organisms with a slow, predictable death (Figure 2A). This pathway towards death resembles to the apoptosis of the cells.

(2) *Cooperation of the network one level higher* may cause the 'over-cooperation' of certain constituent networks. This extensive involvement in the one-level-higher cooperation causes over-perturbation of the constituent, bottom network leading to the exhaustion of bottom network resources. Under these conditions, at the bottom network level damage becomes predominant over repair. Thus over-perturbation may lead to unexpected, stochastic disintegration and rapid death (Figure 2B). This scenario resembles to the necrosis of the cells.

The balance between damage and repair is – obviously – not necessarily unset. We may consider longevity, the healthy ageing of centenarian humans as an example of a well-preserved balance between these two basic factors (Figure 2C). This, successful ageing process is preserving the equilibrium of constituent networks despite their cooperation. However, this is more the unusual exception than the rule. From this point of view the ageing process may be considered as a landscape (the 'ageing-landscape'), where finding the tight trajectory leading to the simultaneous global and local optimum, i.e. the healthy, successful ageing of Figure 2C leading to an exceptionally long life, is similar to find the native state by a protein during the folding process. We may consider the examples of the unbalanced damage and repair of Figure 2A and Figure 2B, as misfolded proteins getting trapped in one or another local minimum of their trajectories.

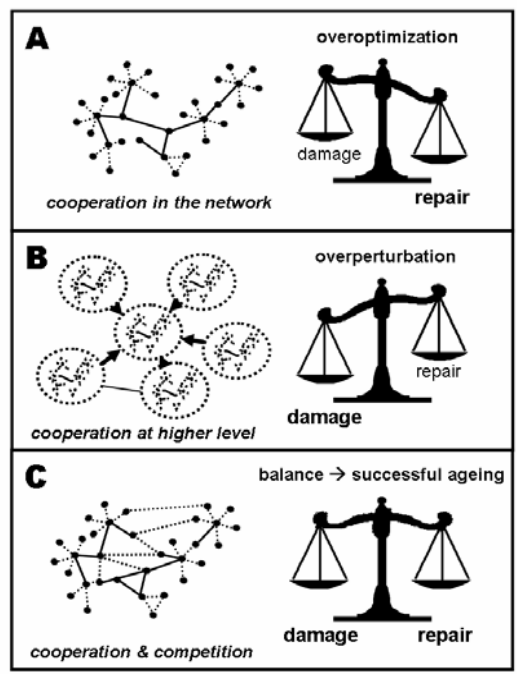

**Figure 2. Ageing as a price of cooperation.** We illustrate the plenitudes of factors setting or un-setting the balance of damage- and repair-related processes. **A.** The equilibrium of damage and repair may be disturbed by over-optimization, where repair becomes predominant over damage. Here the local cooperation of network elements leads to their over-specialization. Network elements form a rigid, locally integrated network typical to aged organisms with a slow, predictable death. **B.** The equilibrium of damage and repair may also be disturbed by over-perturbation. Here over-perturbation comes from 'over-cooperation' of certain networks (like the central network on the left side of the panel) at one level higher. Over-perturbation leads to the exhaustion of resources at the bottom network 'over-cooperating' one level higher. At the bottom network level damage becomes predominant over repair. Over-perturbation may lead to disintegration typical to unexpectedly, stochastically and rapidly dying organisms. **C.** The dynamic balance between cooperation and competition may achieve the balance between damage and repair. This leads to successful ageing, the extreme longevity enjoyed by e.g. the centenarian humans.

Turning back to the more usual unbalanced situations (i.e. the ageing process in the network sense as we defined in this paper) over-optimization of a network may occur, when the environment is stable, and the network has a long, undisturbed time to adapt to a single set of environmental conditions. Under these circumstances networks become rigid and shed off a large segment of their original richness, which had helped them to adapt to the changing environment. A good example of this phenomenon is the reductive evolution of symbiotic organisms, which lose a large segment of their original genome, and form a tightly cooperating metabolic unit inside their host.[63] Another example of network over-optimization is cell differentiation, which leads to cellular senescence. Differentiated cells represent the cooperating small phenotype as opposed by the proliferating, non-cooperating big phenotype of the tumour and stem cell lineages, where cellular senescence never occurs.[64] Over-optimization certainly leads to the excessive loss of symmetry, which was shown to be related to the ageing process. Moreover, both the maintenance of symmetry and the repair of over-optimized cells are costly, which accelerate both the development and ageing of the differentiated, small-phenotype cells during limited resource availability. This connects the over-optimization scenario both to the antagonistic pleiotropy and the disposable soma theories of ageing.[65,66] Network closure, i.e. the development of tightly connected network subsets helps local cooperation, but, when becoming predominant, prevents adaptability and innovation even in social networks.[67] Thus, the over-optimization of network structure to local cooperation develops the rigid, locally integrated network of Figure 1B, which is typical to the 'always-old', slowly ageing organisms.

> **Box 3. Respiration, cooperation and resource-exhaustion: their occurrence in yeasts, tumours, during the evolution of multicellular organisms and consequences in ageing.** Experimental evidence of yeast cells[68] shows that respiratory ATP production yielding about 32 mol of ATP per mole of glucose can be regarded as a form of cooperation. On the contrary, fermentation (having a 16-times lower yield, but a much higher rate of ATP production) is a defective strategy, which uses up the available glucose in an extremely fast manner forcing the competitors to starve.[24] In agreement with the features of competitive and cooperative strategies outlined in Table 1, at a low level of resources cooperation wins over asocial behaviour. However, at a high level of resources non-cooperating, fermenter yeast cells progressively outperform cooperating respirators. Similarly to this, tumour cells most often use fermentation instead of respiration, which is in agreement with their non-cooperating, invasion-prone behaviour.[69] Thus respiration emerges as a pre-requisite of the cooperation necessary to maintain a multicellular organism at high cellular density and low amount of resources. However, respiration (especially respiration-bursts) may lead to the release of free radicals, which promote the ageing of cooperating cells.[70] In agreement with this tumour cells opting for the competitive fermentation do not typically age. Thus the switch from fermentation to respiration is not only a key, cooperating step in the evolution of multicellular, complex systems, but in parallel with this respiration-driven free radicals may exhaust the repair capacity of the host causing an accelerated senescence of the constituent cells of the multicellular organism.

The over-perturbation of a network may occur, when network perturbation and concomitant damage becomes so extensive and continuous, which exhausts the available resources including the repair capacity. An interesting example of resource exhaustion may develop during the switch from fermentation to respiration in yeast at high population densities (Box 3.).[68–70] Here respiration may be regarded as a cooperative strategy, since it does not consume glucose and other energy resources in an inefficient, but fast way as fermentation does. However, the cooperative respiration may lead to an increased production of free radicals, which becomes especially true, when collection efforts of the sparsely available glucose lead to respiration-bursts increasing the level of perturbation further. The low energy resources compromise the repair mechanisms inducing the accumulation of oxidative damage and a consequent ageing.[16,44,71] As an extension of the fermentation/respiration duality, tumour cells are typical competitors: they grow rapidly, most of the times opt for fermentation and do not typically age.[69] We have to emphasize that the above example is true only in the case, when the equilibrium of increased oxidative damage and limited repair capacity becomes severely unbalanced. Moreover, recently more and more examples are published, which question the predominant role of oxidative stress in ageing.[71] This emerging controversy may reflect the dual role of oxidative stress as both a trigger of repair functions, and a continuous burden leading to an overload of the repair mechanisms.

The fermentation to respiration switch of yeast at high population densities was already an example of potential cooperation-induced exhaustion of system resources, where cooperation occurred not at the level of network elements (i.e. proteins of the yeast cell) but between entire networks represented by the yeast cells themselves. Intercellular cooperation may generally compromise intracellular repair mechanisms. As a rather general example there is an increasing overlap between signalling networks and repair functions as we proceed from unicellular organisms to humans. The key signalling elements of the Ras-family, the p53 protein, the poly-ADP-ribose-polymerase, and the critical node of insulin signalling, the Akt/PI-3-kinase complex all overlap with critical repair functions. Conversely, the molecular chaperone families, which provide a central mechanism of protein repair, have numerous connections to various signalling pathways.[60,61]

The repair function at one level higher, at the level of organ-repair is exemplified by wound-repair or bone-repair. These repair functions require the action of the Hedgehog, Notch, TGF-beta, Wnt and growth factor signalling pathways. All these organ-repair signalling pathways have an increasing overlap with the inter-organ communication pathways of the (1) nuclear hormone receptors, (2) JAK-STAT cytokine signalling and (3) insulin as we go from simple multicellular organisms to humans. Here again, a significant suppression of organ-repair capacity may occur by the increase of inter-organ communication during embryo development.[61] The examples can be continued: several layers higher in complexity, much less attention was paid for the repair of brokers' physical and mental health during the recent economic turmoil than at former, 'business as usual' situations. Indeed, during the selling and buying frenzy the intensive cooperation of brokers and clients certainly exhausted their resources (not only financially, but also personally).

Besides the examples shown above, our hypothesis showing that the equilibrium of repair and damage may become unbalanced by (1) cooperation at the network level leading to over-optimized network (Figure 2A), as well as by (2) cooperation at one level higher occasionally over-perturbing some of the constituent networks (Figure 2B) can be tested by a number of additional ways, from which we list a few in Box 4.[26-31,33,36,37,43,59,68,72]

---

**Box 4. Possible tests of the hypothesis.** We suggest the following tests of our hypothesis proposing that (1) cooperation at the network level leading to over-optimized network, as well as (2) cooperation at one level higher occasionally over-perturbing some of the constituent networks both make the equilibrium of repair and damage unbalanced, and both lead to the occurrence of a prolonged damage typical to the ageing process. We have to note that the cooperation of networks scenario is extremely complex, and therefore it is a rather dedicated task to single out cooperation as the only changing parameter during the experimental setup. However, there are still many possibilities to try from which we list only a few.
- **Simulations of cooperating organisms**. In these *in silico* experiments the cooperation of simulated organisms can be studied. If the structure of the cooperating units was planned to be complex enough, the cooperation-induced unbalance, and reduction of fitness can be demonstrated similarly to the studies of ref. 36.
- **'In vitro' evolution studies**. In these studies the cooperation/cheating scenarios of refs. 26-31, 43 or 68 can be extended to examine, if extensive cooperation indeed develops an ageing phenotype.
- **Animal behavioural studies**. These studies partially overlap with the 'in vitro' evolution studies described above, but extend them to more complex organisms. A good example is the study of insect or other animal communities, where cooperation, cheating and their consequences to the ageing process can be examined similarly to refs. 33 and 72.
- **Studies on humans**. Obviously direct studies of this most important subject are difficult to design due to both ethical concerns and time-limitations. However, the extension of studies mentioned in ref. 37 with the ageing and longevity records may give a clue. Insulin resistance has been hypothesized to be an evolutionary conserved cellular process against the damage of reactive oxygen species, which is accelerating the ageing process.[59] A consistent correlation of insulin resistance with the behavioural traits similar to those of ref 37 may also give an additional support of our hypothesis.

---

As we have seen from the above examples, cooperation may occur at all levels of complexity. When the organisms want to catch a lot from the multitude of surrounding resources, they may re-organize their hierarchical networks suppressing local cooperation, while expanding global exploration and successive global cooperation. However, during stress and resource-exhaustion the loose, global cooperation will be inhibited, and the remaining resources become re-channelled to maintain a tight, local cooperation. The former, resource-rich situation may lead to the suppression of repair leading to the sudden ageing of the 'forever-young' network in case of an over-perturbation, while the latter, resource-poor situation may cause the over-optimization of the network, inducing an aged, 'always-old' network structure. However, we have to note that the ageing of the individual networks may not necessary cause the ageing of the network of networks at one level higher of the hierarchical complexity.

Network elements (representing constituent networks themselves) age differently and at different speed. As we described before, non-cooperating, creative elements may preserve the 'forever-young' phenotype even in the middle of an 'always-old' network.[40] Creative entrepreneurs in the age of 50 preserve a risk-taking behaviour typical to the age of 25, which gives a good example of the slow ageing of these non-cooperators.[73] Recent data suggest that the senescence of creative elements, such as inter-modular proteins[74] or stem cells,[19] is an especially important step in the ageing process of the whole complex system. The emerging diversity of either differently over-optimized, or differently damaged networks together with their behavioural diversity help their cooperation.[21,75] This shows that ageing-induced development of diversity and the emergence of cooperation once again, go hand-in-hand, and enhance each other in a successful community.

**Table 3. Cooperation and ageing at major transitions of evolution**

| Evolutionary transition[20] | Types of cooperation* | Cooperation-related ageing | |
|---|---|---|---|
| | | at one level lower in the hierarchy** | at the same level of hierarchy |
| primordial soup → protocell | • macromolecular complexes<br>• metabolic and signalling networks | • accumulated perturbations at hubs and other key points of macromolecular networks leading to increased local molecular damage[40,71] | • growth-induced tensions and subsequent division of the protocell |
| independent replicators → chromosomes | • coordinated RNA/DNA replication | • increased physical constraints leading to a larger probability of RNA/DNA breaks and damage | • chromosomal aberrations and failures |
| RNA-world → genetic code | • coordinated DNA transcription and protein synthesis | • regulation constraints (e.g. limited monomer availability): synthesis of damaged RNA and protein molecules | • transport- and speed-adjustment insufficiencies leading to propagating damage |
| prokaryotes → eukaryotes | • mitochondrial and membrane networks<br>• rearranged, elimination (proteasome)-centred protein-protein interaction network | • mitochondrial free radical production leading to oxidative damage[66] | • energy surplus causing an accelerated, error-prone synthesis of macromolecules and an elimination-centred macromolecule metabolism |
| asexual clones → sexual populations | • mate-selection[72] | • fertility-induced excesses of metabolism and consequent acceleration of ageing | • sexual conflict-induced ageing[73,74] |
| protists → differentiated multicellular organisms | • hormonal regulation<br>• immune system<br>• neuronal networks | • accumulated perturbations at hubs and other key points of cellular networks leading to increased local damage (muscle injuries, immunodeficiencies, neurodegeneration, etc.) | • insufficient mediation of conflict (civilization diseases, autoimmune diseases, etc.) |
| solitary individuals → social networks | • group-related nursing of offspring<br>• division of labour | • increased longevity (menopause)[2] allowing the development of ageing<br>• stress of group-hierarchy establishment | • accelerated dispersal of homogenous network groups[75,76] |
| social networks → human society | • communication and transportation networks<br>• globalization | • information-overload, acceleration-and civilization-induced stress | • environmental pollution (climate-change) |

*In many cases the emerging cooperation helps the elimination of the damage caused by the already established cooperation at one level lower. Thus coordinated replication helps the development of cell division, which is an emerging problem of cell growth; the energy surplus of mitochondria helps the repair of the accumulated damage of concerted transcription and translation, etc. The mediation of conflict in the use of resources is often helped by the complexity of cooperation at one level higher in the organization.
**Ageing at this level does not necessarily lead to the ageing of the next, emerging level of complexity.

**Ageing as consequence of complexity**
In this section we extend our hypothesis outlined in the previous section proposing that cooperation has the additional cost of ageing in a large variety of organisms. In the extension we go beyond the multiple examples of the past section, and show that the above hypothesis can be generalized to all major steps of evolution.[20,58,76–80] All these major steps led to the development of a higher level of complexity, which required the cooperation of more and more complicated parts as detailed in Table 3. Simultaneously with the occurrence of cooperation, in all these evolutionary innovations novel types of cooperation-related constraints have been introduced again and again, which all led to various forms of over-optimization and unbalance between repair and accumulating damage, causing an ageing-like process at the respective level of complexity.

We summarize the most important appearances of evolution-related ageing-type processes in the following examples.
- During the assembly of the protocell, the development of macromolecular complexes and networks led to the appearance of 'hot spots', i.e. macromolecular segments accumulating a high amount of local energy as well as collecting and amplifying the perturbations of the whole, integrated system.[40,58] The amplification of perturbations not only gave excellent chances for increased catalytic actions, but also concentrated the damage and required the emergence of repair mechanisms.
- The evolution of coordinated replication, transcription and protein synthesis induced increasing physical constraints as well as coordination problems of adjustable transport and speed. All of these led to an increase in the local molecular damage as well as in the synthesis of truncated or damaged RNA-s and proteins. The accumulation of such types of damages is a typical sign of the ageing process.
- The development of eukaryotes required the cooperative action of respiration as detailed in Box 3, which set free a continuous bombardment of free radicals, which significantly contribute to the damage-load leading to ageing.
- The development of sexual reproduction induced a variety of sexual conflicts, which have a well-documented contribution to the ageing process.[76–78]
- The appearance of multi-cellular, differentiated organisms not only required a sophisticated and often malfunctioning transport system of nutrients and oxygen, but also led to the damage-inducing amplification of the perturbations at one level higher in the cellular networks of the neurons, immune and muscle cells etc.
- The development of social networks invoked various types of psychosocial stress, which – if experienced in a chronic form – rapidly promote ageing.
- Finally, the current development of globalized communication and transportation networks caused an information-overload, and an acceleration of everyday life, which led to previously inexperienced types of civilization stresses, and caused a massive environmental pollution leading to the currently experienced climate change and the ageing of the global ecosystem.

This chain of events shows that (1) cooperation is a general feature of all major evolutionary innovations at higher and higher levels, and (2) all novel forms of this cooperation evoked novel types of accumulating damage, leading to an ageing-like phenomenon of the respective complex system. Thus, ageing emerges not only as a price of cooperation, but also as a price of the emerging complexity of the self-organizing matter.

**Conclusions: ageing as a price of cooperation and complexity**
In conclusion, in this paper we have outlined the hypothesis that cooperation generally leads to the ageing of cooperating units. First, we gave several examples, where ageing co-occurred with cooperation: (1) we showed that in the two major evolutionary strategies of survival cooperative behaviour is associated with a greater predominance of ageing (Table 1.), (2) we listed the enriched anti-ageing homologues of 'non-cooperative, cheater genes' of the amoeba *Dictyostelium discoideum* (Table 2.) and (3) we described the resemblance of locally cooperative network structures to that of the aged, 'always-old' network (Figure 1).

Next, we outlined an extension of the "network theory of ageing" hypothesis showing that the equilibrium of repair and damage may become unbalanced by (1) cooperation at the network level leading to an over-repaired, over-optimized, 'always-old'-type network structure ageing slowly, and dying in an predictable, apoptosis-like process (Figure 2A). The equilibrium of repair and damage may also become unbalanced by (2) cooperation at one level higher occasionally over-perturbing some of the constituent networks, and exhausting their resources leading to inefficient repair, fast ageing and death in a stochastic, necrosis-like process (Figure 2B). We gave the (a) reductive evolution of symbionts and (b) cell differentiation as examples of over-optimization-induced cellular senescence. We related over-optimization to the antagonistic pleiotropy and disposable soma theories of ageing. To illustrate the multitude of cooperation-related cases leading to resource exhaustion we listed (c) the production of age-promoting free radicals by the respiration-driven cooperative use of external glucose in yeast and differentiated cells (as opposed to fermenting tumours, Box 3), (d) the overlap of signalling pathways with cellular repair mechanisms and (e) the overlap of tissue repair-related and inter-organ communication-related signalling pathways all raising significant conflicts of interest in using the resources of the cells. We also gave examples for over-optimized structures and over-perturbing, resource-exhausting situations in social networks and proposed tests to validate or refute our hypothesis (Box 4).

Finally, we extended the hypothesis and showed that the cooperation of more and more complex units was a necessary behaviour in all major steps of evolution, and it induced a novel, ageing-like phenomenon at each of these evolutionary innovations (Table 3). Thus, ageing emerges as a price of complexity, which is not only induced, but also helped by cooperation going hand-in-hand, and enhancing each other in a successful community.

Our hypothesis not only gives a novel view of ageing by examining the effects of cooperation on the equilibrium of cooperating systems, not only shows in three generally applicable examples that the deterioration of the ageing process co-occurs with cooperation, but gives several examples and test options to show that the equilibrium of repair and damage may become unbalanced by (1) cooperation at the network level leading to an over-optimized network, as well as by (2) cooperation at one level higher over-perturbing some of the constituent networks. Moreover, the hypothesis extends the causative link between cooperation and ageing and proposes that cooperation in all major steps of evolution induced a novel, ageing-like phenomenon. This novel understanding of ageing as a consequence of cooperation gives a novel understanding of the ageing process as a price for the complexity we enjoy on the Earth.

**Acknowledgments** The authors would like to thank the anonymous reviewers, György Szabó (Research Institute for Technical Physics and Material Science; Hungarian Academy of Sciences, Budapest, Hungary) and members of the LINK-Group (www.linkgroup.hu), especially Csaba Böde, Csaba Pál and Balázs Papp for their helpful comments.